\documentclass[twocolumn, showpacs, showkeywords]{revtex4}
\usepackage{epsf}
\usepackage{epsfig}
\usepackage{amsmath}
\usepackage{amssymb}
\usepackage{amsfonts}
\usepackage{dsfont}
\usepackage{dcolumn}
\usepackage{graphicx}
\begin{document}
\draft

\def\VR{{\mathbf r}}
\def\Vt{{\mathbf t}}
\def\VA{{\mathbf a}}
\def\Va{{\mathbf a}}
\def\Vb{{\mathbf b}}
\def\VB{{\mathbf B}}
\def\Vn{{\mathbf n}}
\def\VRR{{\mathbf R}}
\def\VQQ{{\mathbf Q}}
\def\VF{{\mathbf F}}
\def\CF{{\mathcal F}}
\def\Veps{{\mathbf \epsilon}}
\def\Vone{{\mathds 1}}

\title{Fast method for force computations in electronic structure 
calculations}
\author{Nicholas Choly\footnote{Electronic Address: 
    {\tt choly@fas.harvard.edu}} and Efthimios Kaxiras}
\address{Department of Physics and Division of Engineering and Applied Sciences,
         Harvard University,\\
Cambridge, MA 02138}
\date{\today}
\begin{abstract}
  We present new efficient $(O(N \log N))$ methods for computing three 
 quantities crucial to electronic structure calculations: the ionic potential,
 the electron-ion contribution to the Born-Oppenheimer forces, and the
electron-ion contribution to the stress tensor. 
 The present methods are applicable to 
calculations in which the electronic charge density is represented on a
uniform grid in real space.  They are particularly well-suited for metallic
extended systems, where other $O(N)$ methodologies are not readily applicable.
Based on a fast 
 algorithm for determining the atomic structure factor, originally developed 
by Essmann et al. \cite{ESSMANN} for fast Ewald energy and force computation,
the present methods involve approximations that can be systematically
improved.  The methods are tested on a representative metallic system (bulk 
Al), and their ability to simultaneously achieve high accuracy and efficiency
is demonstrated.
\end{abstract}


\maketitle

A wealth of efficient methods have recently been developed
\cite{WYANG, LI, ORDEJON, PENALTY, SIESTA, GOEDECKER, HERNANDEZ, KUDIN,
SHAO, CARTER, YWANG} for calculating
electronic properties of an extended physical system that require an amount
of computation that scales linearly with the size of the system $N$.
This size can be
 defined to be the number of atoms or the number of valence electrons, or
the volume of the system, all of which are linearly related for large
condensed systems.  
In 
this article we present a quasi-linear-scaling $(O(N \log N))$ method for
computing the ionic potential, the ionic forces, and the stress tensor
in electronic structure calculations. 
Atomic forces are necessary for the calculation
of many physical properties of a system, including the determination
of the optimal structure and simulation at a finite temperature.
Some linear-scaling methods achieve linear computational scaling for
the computation of the energy but not for the forces on all of the 
ions\cite{HERNANDEZ}.
Other linear-scaling methods achieve
efficient force calculations by working with a basis of localized
functions\cite{SIESTA, KUDIN, SHAO}, which are less efficient at representing
delocalized electronic states found for example in metallic systems.
The present method 
applies to calculations performed in a periodic parallelepiped supercell,
not necessarily orthogonal,
in which the electronic charge density $\rho(\VR)$ is represented on a 
uniform grid.

As a result of the Hellmann-Feynman\cite{HELLMANN, FEYNMAN} theorem, the 
force on the $p$th
ion (within the Born-Oppenheimer approximation) is given by the sum 
of the partial derivatives of the ion-ion energy (the Ewald energy) and the
electron-ion energy with respect to the atomic coordinates:
\begin{eqnarray}
\VF_{p} = \VF_{p}^{Ewald} + \VF_{p}^{e-i}=-\frac{\partial E^{Ewald}}{\partial
  \Vt_{p}} - \frac{\partial E^{e-i}}{\partial \Vt_{p}}
\end{eqnarray}
where
\begin{eqnarray}
\label{eqn:Epspenergy}
E^{e-i} \equiv \int_{cell} \rho(\VR) V^{ion}(\VR) d\VR  
\end{eqnarray}
and the ionic potential is defined as:
\begin{eqnarray} \label{eqn:vext}
V^{ion}(\VR) = \sum_{\VRR} 
        \sum_{p=1}^{N_{at}} V^{psp}(\VR - 
        \Vt_{p} - \VRR)
\end{eqnarray}
where here the $\Vt_p$ are
the atomic positions within the unit cell; $V^{psp}(r)$ 
is the pseudopotential representing the ions; and the outer sum over $\VRR$
is over all lattice translation vectors $\VRR = n_1 \Va_1 + n_2 \Va_2 +
n_3 \Va_3$ for all integers $n_i$, where the $\Va_i$ are the lattice vectors
defining the unit cell.  For simplicity of
presentation, we will consider systems that involve only one type of 
pseudopotential, but the methods presented here generalize readily to systems
with multiple pseudopotentials.  Furthermore, non-local pseudopotentials
can be split into a short-ranged non-local part and a long-ranged local
part; only the long-ranged local part of $V^{psp}$ will be 
considered here.  
It should be noted that the present methods make only part of the
electronic structure calculation efficient, and in the context of 
the Kohn-Sham method, the overall electronic structure calculation will still
scale as $N^3$ due to the need to orthogonalize $N$ wavefunctions or 
to diagonalize an $N \times N$ matrix. 
Thus the present methods are particularly relevant
to the orbital-free density functional methods\cite{CARTER, YWANG}, which
deal only with local pseudopotentials and can achieve $O(N)$ scaling for the
entire calculation.

The Smooth Particle Mesh Ewald Method\cite{ESSMANN} (SPME) is an efficient 
scheme $(O(N \log N))$
for computing the Ewald energy $E^{Ewald}$ and its derivatives with respect
to the atomic coordinates, $\partial E^{Ewald} / \partial \Vt_{p}$.
Here we show how similar ideas can be used to yield efficient methods 
(also $O(N \log N)$) for
determining the ionic potential $V^{ion}(\VR)$, and the other component
of the Born-Oppenheimer forces, $\partial E^{e-i} / \partial \Vt_{p}$.

We begin presentation of our method by expressing $V^{ion}(\VR)$ in terms
of the structure factor, $S(\VR)$:
\begin{eqnarray}
V^{ion}(\VR)= \frac{1}{\Omega} \int V^{psp}(\VR - \VR') S(\VR') d\VR' 
\end{eqnarray}
where
\begin{eqnarray}
\label{eqn:structfacdef}
S(\VR)=\Omega \sum_{\VRR} \sum_{p=1}^{N_{at}}
  \delta(\VR - \Vt_p - \VRR)
\end{eqnarray}
and $\Omega = |\Va_{1} \cdot (\Va_{2} \times \Va_3)|$ is the unit cell volume.

In reciprocal space, the expression for $V^{ion}$ becomes:
\begin{eqnarray}
\tilde{V}^{ion}_{\VQQ} & \equiv & \frac{1}{\Omega} \int_{cell} V^{ion}(\VR)
        e^{i \VQQ \cdot \VR} d\VR \nonumber \\
\label{eqn:vq}
        &=& \frac{1}{\Omega} \tilde{V}^{psp}(\VQQ) \tilde{S}_{\VQQ}
\end{eqnarray}
where
\begin{eqnarray}
\tilde{V}^{psp}(\VQQ) \equiv \int V^{psp}(\VR) e^{i \VQQ \cdot \VR} d\VR
\end{eqnarray}
and, using Eq. (\ref{eqn:structfacdef}), 
\begin{eqnarray}
\tilde{S}_{\VQQ} &\equiv& \frac{1}{\Omega} \int_{cell} S(\VR) e^{i \VQQ \cdot
        \VR} d\VR \nonumber \\
\label{eqn:structfac}
&=& \sum_{p=1}^{N_{at}} e^{i \VQQ \cdot \Vt_{p}}
\end{eqnarray}
Inversely, we have:
\begin{eqnarray}
V^{ion}(\VR) = \sum_{\VQQ} \tilde{V}^{ion}_{\VQQ}
        e^{-i \VQQ \cdot \VR }
\end{eqnarray}
where the $\VQQ$-sum ranges over all integer multiples of the reciprocal
lattice vectors $\Vb_{i}$ defined by $\VA_{i} \cdot \Vb_j = 2 \pi \delta_{ij}$.

For the purposes of the present work, we assume that $\rho(\VR)$ is 
represented on a $N_1 \times N_2 \times N_3$ grid of points
$\VR_{l_1 l_2 l_3}=\frac{l_1}{N_1}\Va_1 + \frac{l_2}{N_2}\Va_2 + 
        \frac{l_3}{N_3}\Va_3$, with $l_i = 0,...,N_i-1$.  We will denote
quantities such as $\rho(\VR_{l_1 l_2 l_3})$ as  $\rho(l_1, l_2, l_3)$.
The Fourier transform of $\rho(\VR)$ is approximated by:
\begin{eqnarray}
\tilde{\rho}(m_1 \Vb_1 + m_2 \Vb_2 + m_3 \Vb_3)
        \simeq \CF [\rho(l_1,l_2,l_3)]
\end{eqnarray}
where $\CF$ is the discrete Fourier transform:
\begin{eqnarray}
\CF[f(l_1,l_2,l_3)] &=& \frac{1}{N} \sum_{l_1=0}^{N_1-1}
        \sum_{l_2=0}^{N_2-1} \sum_{l_3=0}^{N_3-1} f(l_1,l_2,l_3) \nonumber \\ 
        &\times& e^ {\textstyle 2 \pi i (\frac{l_1 m_1}{N_1} + 
          \frac{l_2 m_2}{N_2} + \frac{l_3 m_3}{N_3}) }
\end{eqnarray}
where $N \equiv N_1 N_2 N_3$.
Its inverse, $\CF^{-1}$, is given by:
\begin{eqnarray}
\CF^{-1}[\tilde{f}(m_1,m_2,m_3)] &=& \sum_{m_1=0}^{N_1-1}
        \sum_{m_2=0}^{N_2-1} \sum_{m_3=0}^{N_3-1} \tilde{f}(m_1,m_2,m_3) 
        \nonumber \\
        &\times& e^{\textstyle -2 \pi i (\frac{l_1 m_1}{N_1} + 
          \frac{l_2 m_2}{N_2} + \frac{l_3 m_3}{N_3})}
\end{eqnarray}
Using Eq. (\ref{eqn:vq}), and transforming back to the real space grid,
we can determine $V^{ion}(\VR)$ via:
\begin{eqnarray}
\label{eqn:Vdiscrete}
\lefteqn{V^{ion}(l_1,l_2,l_3) = } \nonumber \\
&& \frac{1}{\Omega} \CF^{-1}
        \left[ \tilde{P}(m_1,m_2,m_3)\tilde{S}(m_1,m_2,m_3) \right]
\end{eqnarray}
where $\tilde{P}(m_1,m_2,m_3)$ is the array given by:
\begin{eqnarray}
\label{eqn:Pdef}
\tilde{P}(m_1,m_2,m_3) \equiv \tilde{V}^{psp}(m'_1 \Vb_1 + m'_2 
         \Vb_2 + m'_3  \Vb_3)
\end{eqnarray}
and $m'_i=m_i$ for $0 \leq m_i \leq N_i/2$ and $m'_i=m_i-N_i$ otherwise.
The array $\tilde{S}(m_1,m_2,m_3)$ is given by
\begin{eqnarray}
\tilde{S}(m_1,m_2,m_3) \equiv \tilde{S}(m_1 \Vb_1 + m_2  
        \Vb_2 + m_3  \Vb_3)
\end{eqnarray}

Calculation of the structure factor via Eq. (\ref{eqn:structfac}) will scale
with the square of the system size, because it needs to be computed at every
reciprocal space grid point $(m_1,m_2,m_3)$, and at each of these points
 a sum must be performed over
each atom in the system.  However, Essmann et al., as part of their
efficient Smooth Particle Mesh Ewald method\cite{ESSMANN}, provide an
 elegant method for computing the structure factor efficiently (albeit 
approximately),
requiring an  amount of computation that scales as only $N \log{N}$.
By incorporating this algorithm to compute the structure factor,
the present methods for computing both the ionic potential $V^{ion}(\VR)$ 
and the electron-ion contribution to the forces $\VF^{e-i}_{p}$ achieve 
the same $O(N \log N)$ quasi-linear scaling.

Here we summarize the method of Essmann et al. for efficiently computing
the structure factor, but refer the reader to the original reference
\cite{ESSMANN} for full details. 
The crux of the algorithm lies in the approximation of the exponential in 
Eq. (\ref{eqn:structfac}) with (complex) cardinal B-splines.  The $n$th order
cardinal B-spline function,
$M_{n}(x)$, is defined as follows: $M_{2}(x) = 1 - |x-1|$ for
$0 \leq x < 2$, and $M_{2}(x)=0$ otherwise; and the higher-order cardinal
B-splines are defined recursively:
\begin{eqnarray}
M_n(x) = \frac{x}{n-1} M_{n-1}(x)+\frac{n-x}{n-1}M_{n-1}(x-1)
\end{eqnarray}
We define the grid coordinates of the $p$th atom as 
$u_{i p} \equiv N_i \Vt_{p} \cdot \Vb_{i}$, or equivalently
$\Vt_{p} = \frac{u_{1 p}}{N_1}\VA_1 + 
\frac{u_{2 p}}{N_2}\VA_2 + \frac{u_{3 p}}{N_3}\VA_3$. 
The structure factor, expressed in terms of the grid coordinates, is:
\begin{eqnarray}
\label{eqn:gridstructfac}
\lefteqn{ \tilde{S}(m_1,m_2,m_3) = \sum_{p=1}^{N_{at}} \exp \left( 2 \pi i 
\frac{m_1}{N_1}u_{1 p} \right) } \nonumber \\
&& \times \exp \left( 2 \pi i \frac{m_2}{N_2}u_{2 p} \right)
\exp \left( 2 \pi i \frac{m_3}{N_3}u_{3 p} \right)
\end{eqnarray}

The exponentials can be approximated by $n$th order cardinal B-splines, where
$n$ is even, as:
\begin{eqnarray}
\exp \left( 2 \pi i 
\frac{m_j}{N_j}u_{j p} \right) \simeq b_{j}(m_j) \sum_{k=-\infty}
        ^{\infty} M_n (u_{j p} - k) \nonumber \\
        \times \exp \left( 2 \pi i \frac{m_j}{N_j}
         k \right)
\end{eqnarray}
where $b_{j}(m_j)$ is:
\begin{eqnarray}
b_{j}(m_j) &=& \frac{\exp ( 2 \pi i (n-1) m_j / N_j)}
{ \left[\sum\limits_{k=0}^{n-2}
M_n(k+1) \exp (2 \pi i \frac{m_j}{N_j} k) \right] }
\end{eqnarray}
When this B-spline approximation of the exponential is used in 
 Eq. (\ref{eqn:gridstructfac}), it becomes a 
discrete Fourier transform:
\begin{eqnarray}
\lefteqn{ \tilde{S}(m_1,m_2,m_3) \simeq \tilde{B}(m_1,m_2, m_3) } \nonumber \\
&& \times \sum_{p=1}^{N_{at}} 
\sum_{k_1,k_2,k_3=-\infty}^{\infty} 
 M_n (u_{1 p} - k_1) M_n (u_{2 p} - k_2) \nonumber \\
&& \times M_n (u_{3 p} - k_3)
e^{\textstyle 2 \pi i \left(\frac{m_1 k_1}{N_1} + \frac{m_2 k_2}{N_2}+
\frac{m_3 k_3}{N_3} \right) } \nonumber \\
\label{eqn:SBQ}
&& = N \tilde{B}(m_1,m_2,m_3) \CF [Q(l_1,l_2,l_3)]
\end{eqnarray}
where 
\begin{eqnarray}
\label{eqn:Bdef}
\tilde{B}(m_1,m_2,m_3) \equiv b_1(m_1)b_2(m_2) b_3(m_3) 
\end{eqnarray}
and:
\begin{eqnarray}
\label{eqn:qeqn}
Q(l_1,l_2,l_3)= \sum_{p=1}^{N_{at}} \sum_{c_1,c_2,c_3=-\infty}
        ^{\infty} M_n (u_{1 p} - l_1 - c_1 N_1) \nonumber \\
        \times M_n (u_{2 p} - l_2 - c_2 N_2) M_n (u_{3 p} - l_3 - c_3 N_3)
\end{eqnarray}
The array $Q(l_1,l_2,l_3)$ can be computed quickly $(O(N^{at}))$, because 
it is only
non-zero for sub-cubes of dimension $n \times n \times n$ located
near each atom.  It is because $\CF [Q]$ can be computed with the 
fast Fourier transform (FFT) (which is performed in $O(N \log{N})$ operations)
 that the structure factor itself can be computed with $O(N \log N)$ 
operations.

It is now easily seen how the structure factor algorithm provided by
the SPME method can be used to yield an 
efficient method for computing $V^{ion}(l_1,l_2,l_3)$. By substituting
Eq. (\ref{eqn:SBQ}) into Eq. (\ref{eqn:Vdiscrete}), we obtain:
\begin{eqnarray}
\label{eqn:fastVion}
V^{ion}(l_1,l_2,l_3) = \frac{N}{\Omega} \CF^{-1} {\Big [} &&\tilde{P}(m_1,m_2,m_3)
  \tilde{B}(m_1,m_2,m_3) \nonumber  \\
 &&\times  \CF [Q(l_1,l_2,l_3)]  {\Big ]}
\end{eqnarray}
In practice, Eq. (\ref{eqn:fastVion}) could be used to efficiently compute
$V^{ion}(l_1,l_2,l_3)$ with the following algorithm. First $Q(l_1,l_2,l_3)$ 
is computed
via Eq. (\ref{eqn:qeqn}).  Then $\CF [Q]$ is obtained via the FFT.
 $\CF[Q]$ is then multiplied by $\tilde{B}$ and $\tilde{P}$,
defined by Eqs. (\ref{eqn:Bdef}) and (\ref{eqn:Pdef}).  Then the inverse
FFT of this product is computed and multiplied by $N/\Omega$, yielding the
array $V^{ion}(l_1,l_2,l_3)$.

The calculation of the electron-ion contribution to the ionic force, 
$\partial E^{e-i} / \partial \Vt_p$ can also be made efficient, and 
again this comes from expressing $\partial E^{e-i} / \partial \Vt_p$
in terms of the structure factor.
In the discrete variable
representation, the expression for the pseudopotential energy, Eq. 
(\ref{eqn:Epspenergy}), becomes:
\begin{eqnarray}
E^{e-i} \simeq \frac{\Omega}{N} \sum_{l_1, l_2, l_3} \rho(l_1,l_2,l_3) 
        V^{ion}(l_1,l_2,l_3)
\end{eqnarray}
which can be viewed as a dot product of $\rho$ and $V^{ion}$;
we can also evaluate this
dot product in reciprocal space, and taking into account that $\rho$ 
is an array of real numbers, the expression becomes:
\begin{eqnarray}
E^{e-i}  \simeq  \Omega \sum_{m_1,m_2,m_3} \CF \left[\rho(l_1,l_2,l_3) 
        \right]^{\ast}  \CF \left[ V^{ion} (l_1,l_2,l_3) \right]
\end{eqnarray}
then, using Eq. (\ref{eqn:fastVion}) for $V^{ion}$, this becomes:
\begin{eqnarray}
\label{eqn:pisolated}
E^{e-i} & \simeq &  N  \sum_{m_1,m_2,m_3} \CF \left[ \rho(l_1,l_2,l_3)
         \right]^{\ast} \tilde{P}(m_1,m_2,m_3) \nonumber \\
        & & \times \tilde{B}(m_1,m_2,m_3) \CF [Q(l_1,l_2,l_3)]
\end{eqnarray}
Following this substitution, we now express the dot product in real space
again:
\begin{eqnarray}
E^{e-i} & \simeq & \sum_{l_1, l_2, l_3} Q(l_1,l_2,l_3) \CF^{-1} {\Big [}
        \CF \left[\rho(l_1,l_2,l_3) \right] \nonumber \\
         && \times \tilde{P}(m_1,m_2,m_3)^{\ast}
         \tilde{B}(m_1,m_2,m_3)^{\ast} {\Big ]}
\end{eqnarray}
Because the only factor that depends on the atomic positions is $Q$, we
can readily differentiate with respect to the atomic positions:
\begin{eqnarray}
\label{eqn:partial}
\frac{\partial E^{e-i}}{\partial \Vt_{p \alpha}} & = & \sum_{l_1, l_2, l_3} 
        \frac{\partial Q}{\partial \Vt_{p \alpha}} \CF^{-1} {\Big [}
        \CF \left[\rho(l_1,l_2,l_3) \right] \nonumber \\
         && \times \tilde{P}(m_1,m_2,m_3)^{\ast}
         \tilde{B}(m_1,m_2,m_3)^{\ast} {\Big ]}
\end{eqnarray}
where $\alpha = 1,2,3$ is the vector component of the force.  Eqs. 
(\ref{eqn:fastVion}) and (\ref{eqn:partial}) (and the expression for the
stress tensor, Eq. (\ref{eqn:stress})) constitute
the central results of this article.

The partial derivatives $\partial Q / \partial \Vt_{p \alpha}$ can
be evaluated readily with the definition of $Q$, Eq. (\ref{eqn:qeqn}), and
 with the aid of the following B-spline identity:
\begin{eqnarray}
\label{eqn:derivident}
\frac{d}{dx}M_n(x)=M_{n-1}(x)-M_{n-1}(x-1)
\end{eqnarray}

Computing the partial derivatives of Eq. (\ref{eqn:partial}) for all of the
 atoms in the system requires an amount of computation that scales as 
$N \log N $.  As with $Q$, the partial derivative array $\partial Q /
\partial \Vt_{p \alpha}(l_1,l_2,l_3)$ is only non-zero in sub-cubes near
the atomic positions, so computing it requires $O(N)$ computation.  The
fast Fourier transforms scale as $N \log N$.

The application of Eq. (\ref{eqn:partial}) for rapid computation of the
electron-ion forces $\VF^{e-i}$
can proceed algorithmically as follows.
The Fourier transform of
$\rho(l_1,l_2,l_3)$ is computed; then $\CF[\rho]$ is multiplied by 
$\tilde{P}^{\ast}$ and 
$\tilde{B}^{\ast}$. Then the inverse Fourier transform of this product is 
found. Then by utilizing Eq. (\ref{eqn:derivident}), the derivatives 
$\partial Q / \partial \Vt_{p \alpha}(l_1,l_2,l_3)$
are computed during the summing over the $l_i$'s, yielding 
$\partial E^{e-i} / \partial \Vt_{p \alpha}$.

Similar ideas can yield an efficient expression for the computation of the
stress tensor.  The details of the efficient stress tensor method are 
covered in the appendix.

Although these methods have been presented assuming that only one type of
pseudopotential (and hence only one type of ion) is present in the system,
multiple types are readily
treated.  Since the ionic potential is a linear function of the
pseudopotentials, then with multiple types of pseudopotentials the total
ionic potential is a sum of ionic potentials of the different
types, 
\begin{eqnarray}
V^{ion}_{tot}(l_1,l_2,l_3)=\sum_{\tau} V^{ion}_{\tau}(l_1,l_2,l_3)
\end{eqnarray}
where $\tau$ indexes the pseudopotential type.  The individual 
$V^{ion}_{\tau}$ are each computed with Eq. (\ref{eqn:fastVion})
using the $Q$-array associated with this set of ions, and the 
$\tilde{P}$-array associated with this pseudopotential type.
Likewise to compute the ionic forces, Eq. (\ref{eqn:partial}) is used to 
compute the forces on all ions of a given pseudopotential type, using the
$Q$-array associated with this set of ions, and the $\tilde{P}$-array 
associated with this pseudopotential type.

Several tests have been performed to examine the accuracy of these methods.
The accuracy of the approximate structure factor, Eq. (\ref{eqn:SBQ}), 
increases when the number of grid points $N$ increases, and when the B-spline 
order $n$
is increased.  However, in an electronic structure calculation, one is
not simply free to choose the number of grid points for computing $\tilde{S}$.
It is clear from Eqs. (\ref{eqn:fastVion}) and (\ref{eqn:partial}) that
$\tilde{S}$ and the electronic charge density $\rho$ must exist on the
same grid, and typically energy convergence considerations dictate a
minimum grid density on which $\rho$ is represented.
So it must be established that for
a given grid density, the error incurred by using the approximate 
$\tilde{S}$  when  generating $V^{ion}$ and calculating
the forces (i.e. the present methods, Eqs. (\ref{eqn:fastVion}) 
and (\ref{eqn:partial})) is not much
more than the error that would
be present using the same finite number of grid points and the {\em exact}
structure factor $S$ when generating $V^{ion}$ and calculating the forces.
In other words, it must be shown that the error in the total energy and forces
that comes from using an approximate $\tilde{S}$ is smaller than the error
due to using a grid of finite density.
Furthermore, in order to establish that these methods are indeed
(quasi)linear scaling, it must be demonstrated that for a fixed grid density
$N/\Omega$ and fixed B-spline order $n$, the error per atom due to the 
approximate $\tilde{S}$ does not increase when the system size is increased.

All tests here have been done on systems of aluminum atoms simulated with
orbital-free density functional theory\cite{CARTER, YWANG}.  The present 
methods 
for generating $V^{ion}(\VR_i)$ (i.e. the method of Eq. 
(\ref{eqn:fastVion})) and $\VF^{e-i}$ (Eq. (\ref{eqn:partial})) were tested
 as follows:  
\begin{figure}
        \includegraphics[width=0.7\linewidth]{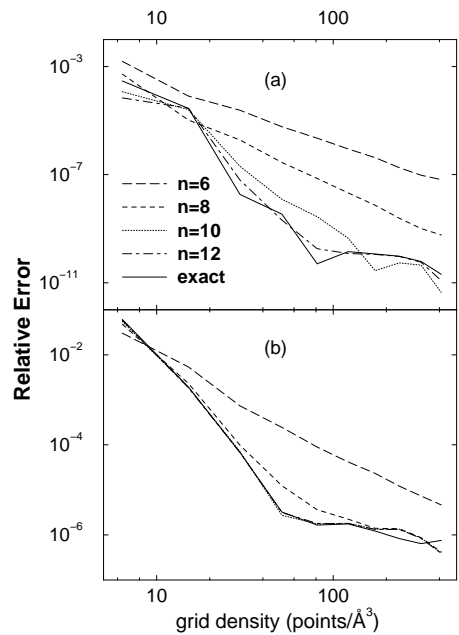}
        \caption{(a) Relative error in the energy compared to infinite
          grid density limit, Eq. (\ref{eqn:energydev}). (b)
          Relative error in $\VF^{e-i}$ compared to the infinite grid
          limit, Eq. (\ref{eqn:forcedev}).}
        \label{fig:vionerror}
\end{figure}
first, in order to test the
error due to the approximate $\tilde{S}$ compared to the error from the
rest of the calculation, 32 Al atoms were placed in a cubic box
 $8.08 {\text \AA}$ on a side, and displaced randomly by about 
$0.5 {\text \AA}$
from their fcc crystalline positions.  Then $V^{ion}$ was generated using the
exact $\tilde{S}$ (Eq. (\ref{eqn:structfac})), and the present method
(Eq. (\ref{eqn:fastVion})) with B-spline orders $n=6,8,10,12$, and separate
electronic relaxations were done in each of these $V^{ion}$'s, yielding
corresponding total energies.  After electronic relaxation, electron-ion 
forces $\VF^{e-i}$ were calculated 
using derivatives of the exact $\tilde{S}$, and with
the present method (Eq. (\ref{eqn:partial})).
This was done for
successively higher grid densities.  The results are shown in 
Fig. \ref{fig:vionerror}.  The relative error between the energy
 and the converged energy, as measured by:
\begin{eqnarray}
\label{eqn:energydev}
\frac{|E-E^{\infty}|}{|E^{\infty}|}
\end{eqnarray}
(where $E^{\infty}$ is the energy in the limit of infinite grid density)
is plotted as a function of grid density for different choices of
$V^{ion}$: $V^{ion}$ generated with the exact method and the
present method.  $E^{\infty}$ is taken to be the total energy evaluated 
with a grid density of 800 points/$\AA^3$, a considerably higher density 
than plotted in Fig. \ref{fig:vionerror}; thus the deviation from this 
grid's energy from the true $E^{\infty}$ is of a smaller order of 
magnitude than the energy deviations found at the grid densities 
explored in Fig. \ref{fig:vionerror}, making it a suitable energy to 
use as $E^{\infty}$.
It is clear that with a B-spline order of $n=10$ the 
error due to the use of the approximate $V^{ion}$ is negligible 
compared to the error due to the finite grid density.  Also plotted is
the relative error in the calculated electron-ion forces $\VF^{e-i}$.  The 
method used for calculating 
the force corresponded to the method used to calculate $V^{ion}$; e.g.
the data for the forces calculated with the present method and a 6th
order B-spline were done with charge densities relaxed in a $V^{ion}$
generated with the present method with a 6th order B-spline.  Thus errors
in the forces that were calculated with the present method have some
error contribution from using the present method for generating $V^{ion}$.
The relative force error was measured as the fractional root-mean-square 
deviation of each force component on each atom:
\begin{eqnarray}
\label{eqn:forcedev}
\left[ \frac{ \sum_{i=1}^{N_{at}} \sum_{\alpha} (F_{i\alpha} - 
  F^{\infty}_{i\alpha})^2}{\sum_{i=1}^{N_{at}} \sum_{\alpha} 
 (F_{i\alpha}^{\infty})^2} \right]^{1/2}
\end{eqnarray}
where the 
$\VF^{\infty}_i$'s are the forces 
 in the limit of infinite grid density, in the same sense as expained
above for $E^{\infty}$.
In this case,
already at a B-spline order of $n=8$ the forces are almost as accurate
as those calculated with the exact method, and for $n=10,12$ the forces
are indistinguishable.

\begin{figure}
        \includegraphics[width=0.7\linewidth]{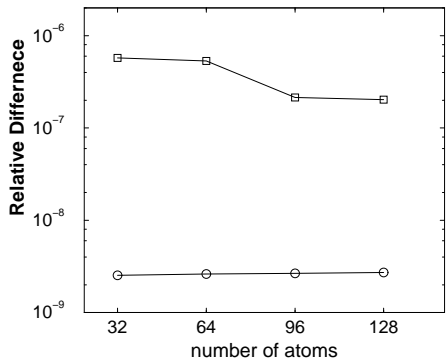}
        \caption{The relative difference between the total energy
          (circles) and forces (squares) calculated with the present
methods and with the exact structure factor methods, as a function
of system size (see Eqs. (\ref{eqn:energydev}) and (\ref{eqn:forcedev})).}
        \label{fig:errorscaling}
\end{figure}

Second, in order to verify the scaling of these methods for different 
system sizes, 
four different systems were considered.  For systems of 32, 64, 96, and 128 
Al atoms, displaced from crystalline positions as before, $V^{ion}$ was
generated with the exact $\tilde{S}$, and with an approximate $\tilde{S}$
generated with 8th-order B-splines. The grid density in each case was 
$237$ points$/{\text \AA}^3$.  After electronic relaxation, forces 
were calculated
using the exact $\tilde{S}$ or with the present method, again corresponding
to the method used to generate $V^{ion}$.  The relative error between these
energies and forces as a function of system size is plotted in 
Fig. \ref{fig:errorscaling}.  The relative errors are measured as before,
with expressions like
Eqs. (\ref{eqn:energydev}) and (\ref{eqn:forcedev}), but instead of 
comparing to the energy and forces of the infinite grid density limit,
the energy and forces are compared to those calculated using the same grid
density and the exact
structure factor of Eq. (\ref{eqn:structfac}).
The relative error in the energy is seen to be constant
as a function of system size, and the relative error in the forces 
is actually seen to decrease slightly with increasing system size.  Thus 
sufficient accuracy can be achieved
with the present methods for calculating $V^{ion}$ and the $\VF^{e-i}$ 
using a fixed B-spline order and grid density, confirming that  these methods
will scale quasi-linearly with system size.

\begin{figure}
  \includegraphics[width=0.7\linewidth]{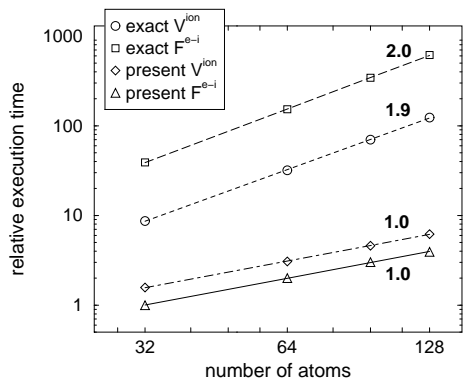}  
  \caption{The computation time required to generate $V^{ion}$ and
    evaluate the $\VF^{e-i}$'s as a function of number of atoms, for 
    the traditional and present methods.  The numbers on each curve
represent the exponent of a power
  law fit.}
  \label{fig:timing}
\end{figure}

Finally, the time required to generate $V^{ion}$
with the exact and present methods, and to evaluate the electron-ion forces
 with the exact and the present methods
is plotted in Fig. \ref{fig:timing}.  One unit of time in the figure 
corresponded to $1.38$ seconds of execution time on a $450$ MHz Pentium III 
processor.
The present methods are far superior
even for modest numbers of atoms.  It is also noted that the present 
methods are readily parallelized.

In conclusion, we present accurate and efficient methods for computing
the ionic potential and the Born-Oppenheimer forces on atoms by 
utilizing an approximate form of
the structure factor.  
It has been demonstrated that the errors due to the present methods do 
not increase with increasing system size.  
As is typically the case with approximate numerical methods,
there is a trade-off between accuracy and efficiency with the present
methods.  Accuracy can be systematically improved by increasing the 
grid density or by increasing the B-spline order $n$, both at the expense 
of more computing time.  However, it was demonstrated that the errors
introduced by using the present methods at a B-spline order of $n=10$ 
are small compared to errors in other parts of the electronic structure 
calculation that arise due to the use of a finite grid density.  Thus
a B-spline order of $10$ is recommended as the optimal compromise for
simple metals like Al, which was used here as a test case.

\appendix
\section{The electron-ion contribution to the stress tensor}
The present methods are readily applied to the computation of the
electron-ion contribution to the stress tensor.  The stress tensor
can be calculated with the methods of Nielsen and 
Martin\cite{NIELSEN}.  One constructs an ionic potential skewed by a 
strain tensor $\Veps$, and a density similarly skewed and
scaled to preserve normalization:
\begin{eqnarray}
V^{ion}_{\epsilon}(\VR) &\equiv& V^{ion} ((\Vone+\Veps)^{-1}\VR), 
\\ \nonumber
\rho_{\epsilon}(\VR) &\equiv& [\det (\Vone + \Veps)]^{-1} \rho 
((\Vone+\Veps)^{-1}\VR)
\end{eqnarray}
The electron-ion contribution to the stress density tensor is then 
given by:
\begin{eqnarray}
\Omega \sigma^{e-i}_{\alpha \beta} =  \left. 
\frac{\partial E^{e-i}_{\epsilon}} {\partial \epsilon_{\alpha \beta}} 
\right|_{\Veps \rightarrow {\mathbf 0}} = \frac{\partial}{\partial
\epsilon_{\alpha \beta}} \int_{cell_\epsilon} \rho_{\epsilon}(\VR) 
V^{ion}_{\epsilon}(\VR) d\VR
\end{eqnarray}
Under the strain transformation, the structure factor is unchanged,
and the Fourier components of the density, $\CF[\rho(l_1,l_2,l_3)]$,
are simply scaled by $[\det (\Vone + \Veps)]^{-1}$.  The reciprocal
lattice vectors, to first order in $\Veps$, become 
$\Vb_i \rightarrow (\Vone - \Veps) \Vb_i$, and hence 
$\partial \Vb_{i \gamma} / \partial \epsilon_{\alpha \beta} =
-\delta_{\alpha \gamma} \Vb_{i \beta}$.  Differentiating Eq. 
(\ref{eqn:pisolated}) with respect to $\epsilon_{\alpha \beta}$, 
one obtains an efficient method for the electron-ion stress density
tensor:
\begin{widetext}
\begin{eqnarray}
\label{eqn:stress}
\sigma^{e-i}_{\alpha \beta} = -\frac{E^{e-i}}{\Omega} 
\delta_{\alpha \beta} -\frac{N}{\Omega} \sum_{m_1,m_2,m_3}
\frac{\VQQ_{\alpha} \VQQ_{\beta}}{|\VQQ|}\tilde{P}'(m_1,m_2,m_3) 
\CF[\rho(l_1,l_2,l_3)]^{\ast} \tilde{B}(m_1,m_2,m_3)
\CF[Q(l_1,l_2,l_3)]
\end{eqnarray}
\end{widetext}

where $\VQQ \equiv m'_1 \Vb_1 + m'_2 \Vb_2 + m'_3 \Vb_3$, and 
$\tilde{P}'(m_1,m_2,m_3) \equiv d \tilde{V}^{psp}(\VQQ) / d |\VQQ|$.
This method has been subjected to simple tests comparing the derivative
of the energy with respect to the crystal lattice constant to the
components of the stress tensor.  These tests indicate an accuracy similar 
to the present force method.  It must be noted, however, that unlike
the ionic forces, the cell stress has contributions from all terms in the
energy, and $\sigma^{e-i}_{\alpha \beta}$
is merely one of them.


\end{document}